\documentclass[a4paper,10pt,twocolumn]{article}
\usepackage{amssymb,amsmath,amsthm}
\usepackage{graphicx}

\begin{document}

\title{Global modeling of transcriptional responses in interaction
  networks\\\small{Leo Lahti\,$^{1}$\footnote{to whom correspondence should
    be addressed}, Juha E.A. Knuuttila\,$^{2}$ and Samuel
  Kaski\,$^{1,*}$\\$^{1}$Aalto University School of Science and
  Technology, Helsinki Institute for Information Technology HIIT and
  Adaptive Informatics Research Centre, Department of Information and
  Computer Science, P.O. Box 15400, FI-00076 Aalto, Finland.
  $^{2}$Neuroscience Center, University of Helsinki, P.O. Box 54,
  FI-00014, Finland}}

\date{}

\maketitle

\def\mub{\boldsymbol{\mu}}
\def\g{\boldsymbol{g}}
\def\xt{\boldsymbol{x}_t}
\def\x{\boldsymbol{x}}
\def\s{\boldsymbol{s}}
\def\Sr{\Sigma_r}
\def\mub{\boldsymbol{\mu}}
\def\g{\boldsymbol{g}}
\def\xt{\boldsymbol{x}_t}
\def\srn{\s_{r}^{(n)}}
\def\Sigrn{\Sigma_{r}^{(n)}}
\def\wrn{w_{r}^{(n)}}
\def\Rn{R^{(n)}}
\def\xn{\x^{(n)}}
\def\rn{r^{(n)}}

\hyphenation{physiologi-cal sub-net-work inter-acti-on tran-script-io-nal associ-ati-ons iden-ti-fi-ca-ti-on characteriza-ti-on}

{\it NOTE: The final version of this manuscript has been published in Bioinformatics 26(21):2713-2720, 2010.}

\begin{abstract}

 {\bf Motivation:} Cell-biological processes are regulated
  through a complex network of interactions between genes and their
  products.  The processes, their activating conditions, and the
  associated transcriptional responses are often
  unknown. Organism-wide modeling of network activation can reveal
  unique and shared mechanisms between physiological conditions, and
  potentially as yet unknown processes.
 {\bf Results:} We introduce a novel approach for organism-wide
  discovery and analysis of transcriptional responses in interaction
  networks. The method searches for local, connected regions in a
  network that exhibit coordinated transcriptional response in a
  subset of conditions. Known interactions between genes are used to
  limit the search space and to guide the analysis.  Validation on a
  human pathway network reveals physiologically coherent responses,
  functional relatedness between physiological conditions, and
  coordinated, context-specific regulation of the genes.
 {\bf Availability:} Implementation is freely available in R and Matlab at\\http://netpro.r-forge.r-project.org/ 
{\bf Contact:} leo.lahti@iki.fi, samuel.kaski@tkk.fi

\end{abstract}

\section{Introduction}

Coordinated activation and inactivation of genes through molecular
interactions determines cell function. Changes in cell-biological
conditions induce changes in the expression levels of co-regulated
genes in order to produce specific physiological responses. A huge
body of information concerning cell-biological processes is available
in public repositories, including gene ontologies \cite{Ashburner00},
pathway models \cite{Schaefer06}, regulatory information
\cite{Loots07}, and protein interactions \cite{Kerrien07}. Less is
known about the contexts in which these processes are activated
\cite{Rachlin06}, and how individual processes are reflected in gene
expression \cite{Montaner09}. Although gene expression measurements
provide only an indirect view to physiological processes, their wide
availability provides a unique resource for investigating gene
co-regulation on a genome- and organism-wide scale. This allows the
detection of transcriptional responses that are shared by multiple
conditions, suggesting shared physiological mechanism with potential
biomedical implications, as demonstrated by the {\it Connectivity map}
\cite{Lamb06} where a number of chemical perturbations on a cancer
cell line were used to reveal shared transcriptional responses between
disparate conditions to enhance screening of therapeutic targets.

\begin{figure*}[ht]
\begin{center}\rotatebox{0}{\resizebox{!}{6.2cm}{\includegraphics{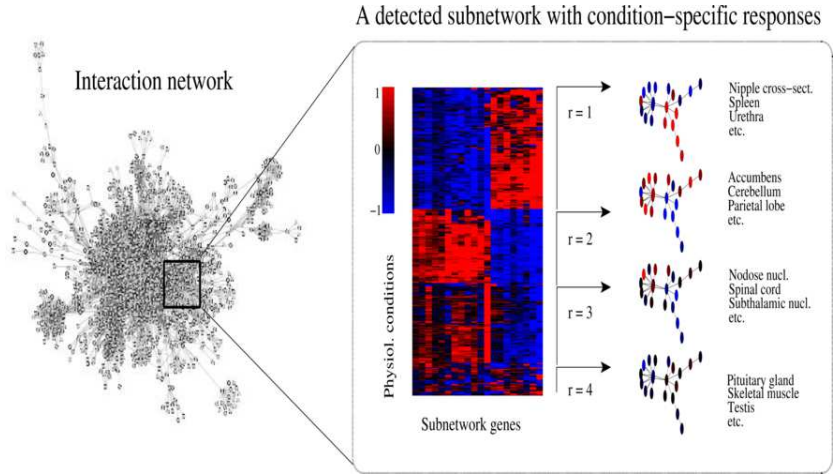}}}\end{center}
\caption{Organism-wide analysis of transcriptional responses in a
  human pathway interaction network reveals physiologically coherent
  activation patterns and condition-specific regulation. One of the
  subnetworks and its condition-specific responses, as detected by the
  NetResponse algorithm is shown. The expression of each gene is
  visualized with respect to its mean level of expression across all
  samples.}
\label{fig:schema}
\end{figure*}

Transcriptional responses have been modeled using so-called {\it gene
  expression signatures} \cite{Hu06}. A signature describes a
co-expression state of the genes, associated with particular
conditions. Well-characterized signatures have proven to be accurate
biomarkers in clinical trials, and hence reliable indicators of cell's
physiological state. Disease-associated signatures are often coherent
across tissues \cite{Dudley09} or platforms \cite{Hu06}.  Commercial
signatures are available for routine clinical practice
\cite{Nuyten08}, and other applications have been suggested recently
\cite{Dudley09}. The established signatures are typically designed to
provide optimal classification performance between two particular
conditions. The problem with the classification-based signatures is
that their associations to the underlying physiological processes are
not well understood \cite{Lucas09}. Our goal is to enhance the
understanding by deriving transcriptional signatures that are
explicitly connected to well-characterized processes through the
network.

We introduce and validate a novel approach for organism-wide discovery
and analysis of transcriptional response patterns in interaction
networks. Our algorithm has been designed to detect and model local
regions in a network, each of which exhibits similar transcriptional
response in a subset of conditions.  The algorithm is independent of
predefined classifications for genes or conditions. This extends the
previous network-based approaches that detect differentially expressed
subnetworks between two predefined conditions \cite{Ideker02,
  Sanguinetti08}. Organism-wide analysis can reveal unique and shared
mechanisms between disparate conditions \cite{Lage08}, and
potentially as yet unknown processes \cite{Nacu07}. The proposed
NetResponse algorithm provides an efficient model-based tool for
simultaneous feature selection and class discovery that utilizes known
interactions between genes to guide the analysis.  Related approaches
include cMonkey \cite{Reiss06} and a modified version of SAMBA
biclustering \cite{Tanay04}. These are application-oriented tools
that rely on additional, organism-specific information, and their
implementation is currently not available for most organisms,
including human. We provide a general-purpose algorithm whose
applicability is not limited to particular organisms.

NetResponse makes it possible to perform data-driven identification of
functionally coherent network components and their condition-specific
responses.  This is useful since the commonly used alternatives,
predefined gene sets or pathways, are collections of intertwined
processes rather than coherent functional entities
\cite{Nacu07}. This has complicated their use in gene expression
analysis, and methods have consequently been suggested for identifying
the 'key condition-responsive genes' of predefined gene sets
\cite{Lee08c}, or for decomposing predefined pathways into smaller
functional modules represented by gene expression signatures
\cite{Chang09}. Our network-based search procedure detects the
coordinately regulated gene sets in a data-driven manner.  Gene
expression provides functional information of the network that is
missing in purely graph-oriented approaches for studying
cell-biological networks \cite{Aittokallio06}.  The network brings in
prior information of gene function and connects the responses
more closely to known processes. This would be missing from purely
gene expression-based methods such as biclustering \cite{Madeira04},
subspace clustering, or other feature selection approaches
\cite{Law04, Roth04}. A key difference to previous network-based
clustering methods, including MATISSE \cite{Ulitsky07} and related
approaches \cite{Hanisch02, Shiga07} is that they assume a single
correlated reponse between all genes in a module. NetResponse
additionally models condition-specific responses of the network. This
allows a more expressive definition of a functional module, or a
signature.

We validate the algorithm by modeling condition-specific
transcriptional responses in a human pathway interaction network
across an organism-wide collection of physiological conditions. The
results highlight functional relatedness between tissues, providing a
global view on cell-biological network activation patterns.

\section{Methods}

\subsection{The NetResponse Algorithm}

We introduce a new approach for global detection and characterization
of transcriptional responses in genome-scale interaction
networks. NetResponse searches for local, connected {\it subnetworks}
where joint modeling of gene expression reveals coordinated
transcriptional response in particular conditions
(Fig.~1). More generally, it is a new algorithm for
simultaneous feature selection (for genes) and class discovery (for
conditions) that utilizes known interactions between genes to limit
the search space and to guide the analysis.

\subsubsection*{Gene expression signatures.}

Subnetworks are the functional units of the interaction network in our
model; transcriptional responses are described in terms of subnetwork
activation. Given a physiological state, the underlying assumption is
that gene expression in subnetwork \(n\) is regulated at particular
levels to ensure proper functioning of the relevant processes. This
can involve simultaneous activation and repression of the genes:
sufficient amounts of mRNA for key proteins has to be available while
interfering genes may need to be silenced. This regulation is reflected
in a unique expression signature \(\s^{(n)}\), a vector describing the
associated expression levels of the subnetwork genes.  The level of
regulation varies from gene to gene; expression of some genes is
regulated at precise levels whereas other genes fluctuate more
freely. Given the physiological state, we assume that the distribution of
observed gene expression is Gaussian, \(\xn \sim N(\s^{(n)},
\Sigma^{(n)})\).

\subsubsection*{Modeling condition-specific transcriptional responses.}

Each subnetwork is potentially associated with alternative
transcriptional states, activated in different conditions and
corresponding to unique combinations of processes.  Since individual
processes and their transcriptional responses are in general unknown
\cite{Lee08c}, detection of condition-specific responses provides an
efficient proxy for identifying functionally distinct states of the
network. Our task is to detect and characterize these signatures.  We
assume that in a specific observation (measurement condition), the
subnetwork $n$ can be in any one of \(\Rn\) latent physiological
states indexed by $r$. Each state is associated with a unique
expression signature \(\srn\) over the subnetwork genes.  Associations
between the observations and the underlying physiological states are
unknown, and treated as latent variables. This leads to a mixture
model for gene expression in the subnetwork \(n\):
\begin{equation}
  \xn \sim \sum_{r=1}^{\Rn} \wrn p(\xn | \srn, \Sigrn),
  \label{eq:gm}
\end{equation}
where each component distribution $p$ is assumed to be Gaussian.
In practice, we assume a diagonal covariance matrix \(\Sigrn\).

A particular transcriptional response is characterized by the triple
\(\{\srn, \Sigrn, \wrn\}\). This defines the shape, fluctuations, and
frequency of the associated gene expression signature in subnetwork
\(n\). The feasibility of the Gaussian modeling assumption is
supported by the previous observations of \cite{Kong06}, where
predefined gene sets were used to investigate differences in gene
expression between two predefined sample groups. In our model, the
subnetworks, transcriptional responses and the activating conditions
are learned from data. In one-channel data such as Affymetrix arrays
used in this study, the centroids \(\srn\) describe absolute
expression signals of the preprocessed array data. Relative
differences can be investigated by comparing the detected responses.
The model is applicable also on two-channel expression data when a
common reference sample is used for all arrays since the relative
differences are not altered by the choice of comparison baseline when
the same baseline is used for all samples.

Now the model has been specified assuming the subnetworks are given.
In practice they are learned from the data. In order to do this we
make two assumptions. First, we rely on the prior information in the
global interaction network, and assume that co-regulated gene groups
are connected components in this network. Second, we assume that the
subnetworks are independent. This allows a well-defined algorithm, and
the subnetworks are then interpretable as independent components of
transcriptional regulation.  In practice the algorithm, described
below, is an agglomerative approximation for searching for locally
independent subnetworks.

\begin{figure}[ht!]
\begin{center}\rotatebox{0}{\resizebox{!}{2.6cm}{\includegraphics{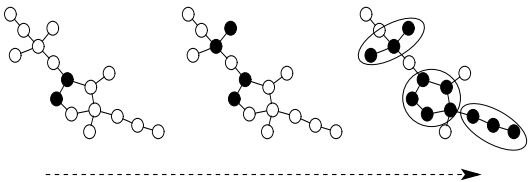}}}\end{center}
\caption{The agglomerative subnetwork detection procedure. Initially,
  each gene is assigned in its own singleton subnetwork. Agglomeration
  proceeds by at each step merging the two neighboring subnetworks
  that benefit most from joint modeling of their transcriptional
  responses. This continues until no improvement is obtained by
  merging the subnetworks.}
\label{fig:agglomeration}
\end{figure}

\subsection{Implementation}\label{sec:implementation}

Efficient implementation is crucial for scalability.  For fast
computation, we use an agglomerative procedure where interacting genes
are gradually merged into larger subnetworks (Fig.~2).  Joint modeling
of dependent genes reveals coordinated responses and improves the
likelihood of the data when compared to independent models, giving the
first criterion for merging the subnetworks. However, increasing
subnetwork size tends to increase model complexity and the possibility
of overfitting since the number of samples remains constant while the
dimensionality (subnetwork size) increases. To compensate for this
effect, we use a Bayesian information criterion \cite{Gelman03} to
penalize increasing model complexity and to determine optimal
subnetwork size.

The cost function for a subnetwork \(G\) is \(C(G) = - 2L + q
Log(N)\), where \(L\) is the (marginal) log-likelihood of the data,
given the mixture model in Eq.~\ref{eq:gm}, \(q\) is the the number of
parameters, and \(N\) denotes sample size. NetResponse searches for a
joint model for the network genes that maximizes the likelihood of
observed gene expression, but avoids increasing model complexity
through penalizing an increasing number of model parameters. An
optimal model is searched for by at each step merging the subnetwork
pair that produces the maximal gain in the cost function. More
formally, the algorithm merges at each step the subnetwork pair \(G_i,
G_j\) that minimizes the cost \(\Delta\mathcal{C} = - 2(L_{i,j} - (L_i +
L_j)) + (q_{i,j} - (q_i + q_j))Log(N) \).  The agglomerative scheme is
as follows:

{\it Initialize:} Learn univariate Gaussian mixture for the expression
values of each gene, and bivariate joint models for all potential gene
pairs with a direct link. Assign each gene into its own singleton
subnetwork.

{\it Merge:} Merge the neighboring subnetworks \(G_i\), \(G_j\) that
have a direct link in the network and minimize the difference
\(\mathcal{C}\). Compute new joint models between the newly merged
subnetwork and its neighbors.

{\it Terminate:} Continue merging until no improvement is obtained by
merging the subnetworks (\(\Delta\mathcal{C} \geq 0\)).

The number \(\Rn\) of distinct transcriptional responses of the
subnetwork is unknown, and is estimated with an infinite mixture
model. Learning several multivariate Gaussian mixtures between the
neighboring subnetworks at each step is a computationally demanding
task, in particular when the number of mixture components is
unknown. The Gaussian mixtures, including the number of mixture
components, are learned with an efficient variational Dirichlet
process implementation \cite{Kurihara07nips}.  The likelihood \(L\)
in the model is approximated by the lower bound of the variational
approximation.  The Gaussian mixture detects a particular type of
dependency between the genes. In contrast to MATISSE \cite{Ulitsky07}
and other studies that use correlation or other methods to measure
global co-variation, the mixture model detects coordinated responses
that can be activated only in a few conditions. Condition-specific
joint regulation indicates functional dependency between the genes but it
may have a minor contribution to the overall correlation between gene
expression profiles. In principle, we could also model the
dependencies in gene fluctuations within each individual response with
covariances of the Gaussian components. However, this would heavily
increase model complexity, and therefore we leave dependencies in
gene-specific fluctuations within each response unmodeled, and focus
on modeling differences between the responses. NetResponse provides a
full generative model for gene expression, where each subnetwork is
described with an independent joint mixture model.  The maximum
subnetwork size is limited to 20 genes to avoid numerical
instabilities in computation. The infinite Gaussian mixture can
automatically adapt model complexity to the sample size. We model
subnetworks of 1-20 genes across 353 samples; similar dimensionality
per sample size has previously been used with variational mixture
models \cite{Honkela08}.

\subsection{Data}\label{sec:data}

\subsubsection*{Pathway interaction network.}

We investigate the pathway interaction network based on the KEGG
database of metabolic pathways \cite{Kanehisa08} provided by the SPIA
package \cite{Tarca09} of BioConductor (www.bioconductor.org).  This
implements the pathway impact analysis method originally proposed in
\cite{Draghici07}, which is to our knowledge currently the only
freely available pathway analysis tool that considers pathway
topology. SPIA provides the data in a readily suitable form for our
analysis. Other pathway data sets, commonly provided in the BioPAX
format, are not readily available in a suitable pairwise interaction
form. Directionality and types of the interactions were not
considered.  Genes with no expression measurements were removed from
the analysis. We investigate the largest connected component of the
network with 1800 unique genes, identified by Entrez GeneIDs.

\subsubsection*{Gene expression data.}

We analyzed a collection of normal human tissue samples from ten
post-mortem donors \cite{Roth06}, containing gene expression
measurements from 65 normal physiological conditions.  To ensure
sample quality, RNA degradation was minimized in the original study by
flash freezing all samples within 8.5 h postmortem. Only the samples
passing Affymetrix quality measures were included. Each condition has
3-9 biological replicates measured on the Affymetrix HG-U133plus2.0
platform. The reproducibility of our findings is investigated in an
independent human gene expression atlas \cite{Su04}, measured on the
Affymetrix HG-U133A platform, where two biological replicates are
available for each measured condition.  In the comparisons we use the
25 conditions available in both data sets (adrenal gland cortex,
amygdala, bone marrow, cerebellum, dorsal root ganglia, hypothalamus,
liver, lung, lymph nodes, occipital lobe, ovary, parietal lobe,
pituitary gland, prostate gland, salivary gland, skeletal muscle,
spinal cord, subthalamic nucleus, temporal lobe, testes, thalamus,
thyroid gland, tonsil, trachea, and trigeminal ganglia).  Both data
sets were preprocessed with RMA \cite{Irizarry03rma}.  Certain genes
have multiple probesets, and a standard approach to summarize
information across multiple probesets is to use alternative probeset
definitions based on probe-genome remapping \cite{Dai05}. This would
provide a single expression measure for each gene. However, since the
HG-U133A array represents a subset of probesets on the HG-U133Plus2.0
array, the redefined probesets are not technically identical between
the compared data sets. To minimize technical bias in the comparisons,
we use probesets that are available on both platforms. Therefore, we
rely on manufacturer annotations of the probesets and use an
alternative approach (used e.g. by \cite{Nymark07}), where one of the
available probesets is selected at random to represent each unique
gene. Random selection is used to avoid selection bias. When
available, the 'xxxxxx\_at' probesets were used because they are more
specific by design than the other probe set types
(www.affymetrix.com).

\subsection{Validation}\label{sec:validation}

The NetResponse algorithm is validated with an application on the
pathway interaction network of 1800 genes \cite{Tarca09} across 353
gene expression samples from 65 physiological conditions in normal
human body \cite{Roth06}. NetResponse is compared to alternative
approaches in terms of physiological coherence and reproducibility of
the findings.

\subsubsection*{Comparison methods.}

NetResponse is designed for organism-wide modeling of transcriptional
responses in genome-scale interaction networks. Simultaneous detection
of the subnetworks and their condition-specific responses is a key
feature of the model. A straightforward alternative would be a
two-step approach where the subnetworks and their condition-specific
responses are detected in separate steps, although this can be
unoptimal for detecting condition-specific responses. Various methods
are available for detecting subnetworks based on network and gene
expression data \cite{Hanisch02, Shiga07} in the two-step
approach. We use MATISSE, a state-of-the-art algorithm described in
\cite{Ulitsky07}.  MATISSE finds connected subgraphs in the network
such that each subgraph consists of highly correlated genes. The
output is a list of genes for each detected subnetwork. Since MATISSE
only clusters the genes, we model transcriptional responses of the
detected subnetworks in a separate step by using a similar mixture
model to the NetResponse algorithm. This combination is also new, and
called MATISSE+ in this paper. The second comparison method is the
SAMBA biclustering algorithm \cite{Tanay02bioinf}. The output is a
list of associated genes and conditions for each identified
bicluster. SAMBA detects gene sets with condition-specific responses
but, unlike NetResponse and MATISSE+, the algorithm does not utilize
the network. Influence of the prior network is additionally
investigated by randomly shuffling the gene expression vectors, while
keeping the network and the within-gene associations intact.
Comparisons between the original and shuffled data help to assess
relative influence of the prior network on the results. Comparisons to
randomly shuffled genes in SAMBA are not included since SAMBA does not
use the network.

\subsubsection*{Reproducibility in validation data.}

Reproducibility of the findings is investigated in an independent
validation data set in terms of significance and correlation (for
details, see Section~\ref{sec:data}).  Each comparison method
implies a grouping for the physiological conditions in each
subnetwork, corresponding to the detected responses. It is expected
that physiologically relevant differences between the groups are
reproducible in other data sets. We tested this by estimating
differential expression between the corresponding conditions in the
validation data for each pairwise comparison of the predicted groups
using a standard test for gene set analysis (GlobalTest;
\cite{Goeman04}). To ensure that the responses are also qualitatively
similar in the validation data, we measured Pearson correlation
between the detected responses and those observed in the corresponding
conditions in validation data.  The responses were characterized by
the centroids provided by the model in NetResponse and MATISSE+.  For
SAMBA we used the mean expression level of each gene within each group
of conditions since SAMBA groups the conditions but does not
characterize the responses. In validation data, the mean expression
level of each gene is used to characterize the response within each
group of conditions. Probesets were available for 75\% of the genes in
the detected subnetworks in the validation data; transcriptional
responses with less than three probesets in the validation data were
not considered. Validation data contained corresponding samples for
\(>79\%\) of the predicted responses in NetResponse, MATISSE+, and
SAMBA (Supplementary Table~1).

\section{Results}

The validation results reported below demonstrate that the NetResponse
algorithm is readily applicable for modeling transcriptional responses
in interaction networks on an organism-wide scale.  While biomedical
implications of the findings require further investigation,
NetResponse detects a number of physiologically coherent and
reproducible transcriptional responses in the network, and highlights
functional relatedness between physiological conditions. It also
outperformed the comparison methods in terms of reproducibility of the
findings.

\subsection{Application to human pathway network}

In total, NetResponse identified 106 subnetworks with 3-20 genes
(Supplementary data file). For each subnetwork, typically (median) 3
distinct transcriptional responses were detected across the 65
physiological conditions (Supplementary Fig.~1). One of the
subnetworks with four distinct responses is illustrated in
Fig.~1.  Each respose is associated with a subset of
conditions. Statistically significant differences between the
corresponding conditions were observed also in the independent
validation data (\(p < 0.01\); GlobalTest). Three of the four
responses were also qualitatively similar (correlation \(>0.8\);
Supplementary Fig.~2). The first response is associated with
immune-system related conditions such as spleen and tonsil. Responses
2-3 are associated with neuronal conditions such as subthalamic or
nodose nucleus, or with central nervous system, for example accumbens
and cerebellum. The fourth group manifests a 'baseline' signature that
fluctuates around the mean expression level of the genes. Testis and
pituitary gland are examples of conditions in this group. While most
physiological conditions are strongly associated with a particular
response, samples from amygdala, bone marrow, cerebral cortex, heart
atrium, and temporal lobe manifested multiple responses.  In general,
it is not well known how individual pathways are manifested at gene
expression level. While alternative responses reveal
condition-specific regulation, detection of physiologically coherent
and reproducible responses may indicate shared mechanisms between
physiological conditions.  Although the responses may reflect
previously unknown processes, it is likely that some of them reflect
the activation patterns of known pathways.  Overlapping pathways can
provide a starting point for interpretation. The subnetwork of
Fig.~1 overlaps with various known pathways, most
remarkably with the MAPK pathway with 10 genes (detailed gene-pathway
associations are provided in the Supplementary data file; see
subnetwork 12). MAPK is a general signal transduction system that
participates in a complex, cross-regulated signaling network that is
sensitive to cellular stimuli \cite{Wilkinson00}. Association of MAPK
to cell growth and proliferation could potentially explain the
differences between neuronal and other conditions. Six subnetwork
genes participate in the p53 pathway, which is a known regulator of
the MAPK signaling pathway. In addition, p53 is known to interact with
a number of other pathways, both as an upstream regulator, and a
downstream target \cite{Wu04a}.  Both MAPK and p53 are associated
with processes including cell growth, differentiation, and apoptosis,
and exhibit diverse cellular responses to varying
conditions. Condition-specific regulation can potentially explain the
detection of alternative transcriptional states of the subnetwork.

The detected responses characterize absolute expression signals in our
preprocessed one-channel array data. Systematic differences in the
expression levels of the individual genes are normalized out in the
visualization by showing the relative expression of each gene with
respect to its mean expression level across all samples. Note that the
choice of a common baseline does not affect the relative differences
between the samples.

\begin{figure}[ht]
\vspace{-5mm}
\rotatebox{0}{\includegraphics[width=.5\textwidth]{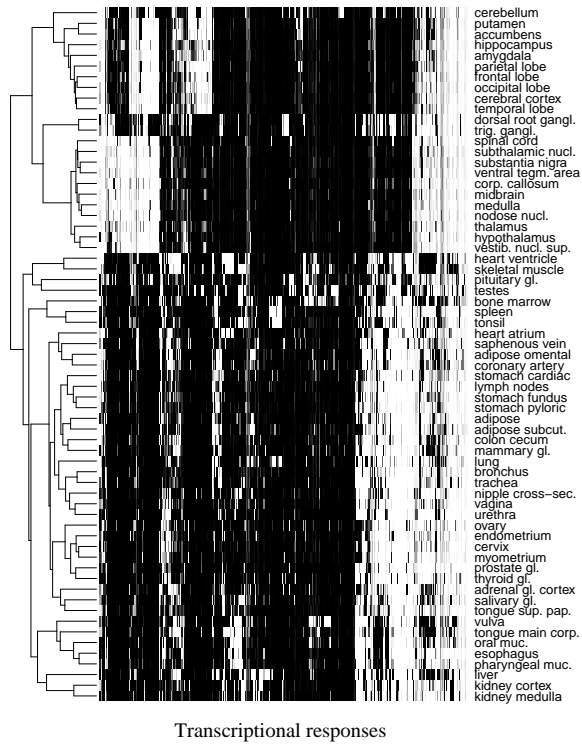}}
\vspace{-7mm}
\caption{Associations between 65 physiological conditions (rows) and the
  detected transcriptional responses of the pathway interaction
  network of Fig.~1. The shade indicates the
  probability of a particular transcriptional response in each
  condition (black: \(P=0\); white: \(P=1\)). Hierarchical clustering
  based on the signature co-occurrence probabilities between each pair
  of physiological conditions highlights their relatedness.}
\label{fig:pcoh}
\end{figure}

\subsubsection*{Condition-selective network activation.}

Associations between the physiological conditions and the detected
transcriptional responses are shown in Fig.~3. Some responses are
shared by many conditions, while others are more specific to
particular contexts such as immune system, muscle, or the
brain. Related physiological conditions often exhibit similar network
activation patterns, which is seen by grouping the conditions
according to co-occurrence probabilities of shared transcriptional
response.  This is known as {\it tissue-selectivity} of gene
expression \cite{Liang06}.

\subsubsection*{Probabilistic tissue connectome.}

Relatedness of physiological conditions can be measured in terms of
shared transcriptional responses (Supplementary Fig.~3).  This is an
alternative formulation of the {\it tissue connectome} map suggested
by \cite{Greco08} to highlight functional {\it connectivity} between
tissues based on the number of shared differentially expressed genes
at different thresholds. We use shared network responses instead of
shared gene count. The use of co-regulated gene groups is expected to
be more robust to noise than the use of individual genes. As the
overall measure of connectivity between physiological conditions, we
use the mean of signature co-occurrence probabilities over the
subnetworks, given the model in Eq.~\ref{eq:gm}. The analysis reveals
functional relatedness between the conditions. In particular, two
subcategories of the central nervous system appear distinct from the
other conditions. Closer investigation of the observed responses would
reveal how the conditions are related at transcriptional level
(Supplementary data file).

\begin{figure}[t]
\centerline{\rotatebox{0}{\includegraphics[width=.5\textwidth]{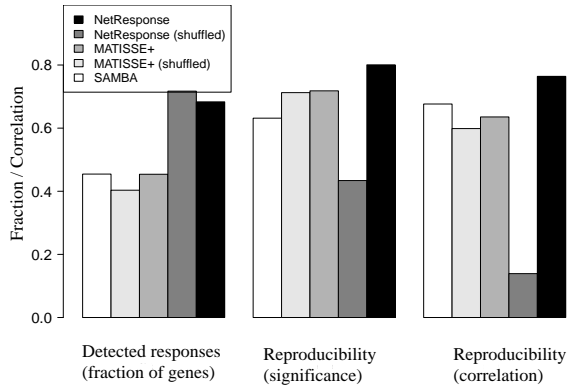}}}
\caption{Comparison between the alternative approaches. {\it Detected
    responses:} Fraction of genes participating in the detected
  transcriptional responses.  {\it Reproducibility (significance):}
  Fraction of responses that are reproducible in the validation data
  in terms of differential expression between the associated
  conditions (\(p<0.05\); GlobalTest).  {\it Reproducibility
    (correlation):} Median correlation between the gene expression
  levels of the detected responses and the corresponding conditions in
  the validation data.}
\label{fig:comparisons}
\end{figure}

\subsection{Comparison to alternative approaches}

NetResponse was compared to the alternative approaches in terms of
physiological coherence and reproducibility of the findings
(Fig.~4; Supplementary Table~1). NetResponse
detected the largest amount of responses; 68\% of the network genes
were associated with a response, compared to 45\% in MATISSE+ and
SAMBA. At the same time, NetResponse outperformed the comparison
methods in terms of reproducibility of the findings.

\subsubsection*{Physiological coherence.}

The association between the responses and physiological conditions was
measured by normalized mutual information (NMI; \cite{Bush08}) between
the sample-response assignments and sample class labels within each
subnetwork. The NMI varies from 0 (no association) to 1 (deterministic
association).  The transcriptional responses detected by NetResponse,
MATISSE+, and SAMBA show statistically significant associations to
particular physiological conditions with a significantly higher
average NMI (0.46-0.50) than expected based on randomly labeled data
(0.26-0.32; \(p< 10^{-4}\); Wilcoxon test; Supplementary Table~1). The
highest average NMI (0.50) was obtained by NetResponse but differences
between NetResponse, MATISSE+, and SAMBA are not significant.
NetResponse is significantly physiologically more coherent also when
compared to results obtained with shuffled gene expression (NMI 0.22;
\(p< 10^{-12}\)). The observations confirm the potential physiological
relevance of the findings in NetResponse, MATISSE+, and SAMBA.

\subsubsection*{Reproducibility.}

The majority of the detected responses were reproducible both in terms
of significance and correlation (Supplementary Fig. 4) as described in
Section~\ref{sec:validation}.  Of the predicted differences between
groups of physiological conditions, 80\% were significant in
validation data with \(p<0.05\) (GlobalTest), compared to 72\% and
63\% in MATISSE+ and SAMBA, respectively, or 43\% obtained for
randomly shuffled data with NetResponse (Fig.~4). The changes were
also qualitatively similar; in NetResponse the median correlation
between the detected responses and corresponding conditions in the
validation data is 0.76, which is significantly higher (\(p<0.01\);
Wilcoxon test) than in the comparison methods (MATISSE+: 0.64; SAMBA:
0.68), or in randomly shuffled NetResponse data (0.14). NetResponse
detected responses for a larger fraction of the genes (68\%) than the
other methods.  This seems an intrinsic property of the algorithm
since it detected responses for a similar fraction of the genes also
in the network with randomly shuffled genes (72\%). However, only the
findings from the real data were reproducible.

\section*{Discussion}

Cell-biological networks may cover thousands of genes, but any change
in the physiological context typically affects only a small part of
the network. While gene function and interactions are often subject to
condition-specific regulation \cite{Liang06}, they are typically
studied only in particular experimental conditions.  Organism-wide
analysis could reveal highly specialized functions that are activated
only in one or a few conditions. Detection of shared responses between
the conditions can reveal previously unknown functional connections
and help to formulate novel hypotheses of gene function in previously
unexplored contexts. We provide a well-defined algorithm for such
analysis.

The results support the validity of the model. NetResponse detected
the largest number of responses without compromising physiological
coherence or reproducibility of the findings compared to the
alternatives. The most highly reproducible results were obtained by
NetResponse. Further analysis is needed to establish the physiological
role of the findings.

NetResponse is readily applicable for modeling condition-specific
responses in cell-biological networks, including pathways, protein
interactions, and regulatory networks. The network connects the
responses to well-characterized processes, and provides readily
interpretable results that are less biased towards known biological
phenomena than methods based on predefined gene sets that are
routinely used in gene expression studies to bring in prior
information of gene function and to increase statistical
power. However, these are often collections of intertwined processes
rather than coherent functional entities. For example, pathways from
KEGG may contain hundreds of genes while only a small part of a
pathway may be affected by changes in physiological conditions
\cite{Nacu07}.  This has complicated the use of predefined gene sets
in gene expression studies. \cite{Draghici07} demonstrated that taking
into account aspects of pathway topology, such as gene and interaction
types can improve the estimation of pathway activity. While their SPIA
algorithm measures the activity of known pathways between two
predefined conditions, our algorithm searches for potentially unknown
functional modules, and detects their association to multiple
conditions simultaneously. This is useful since biomedical pathways
are human-made descriptions of cellular processes, often consisting of
smaller, partially independent modules \cite{Chang09,
  Hartwell99}. Our data-driven search procedure can rigorously
identify functionally coherent network modules where the interacting
genes show coordinated responses.  Joint modeling increases
statistical power which is useful since gene expression, and many
interaction data types such as protein-protein interactions, have high
noise levels. The probabilistic formulation accounts for biological
and measurement noise in a principled manner. Certain types of
interaction data such as transcription factor binding or protein
interactions are directly based on measurements. This can potentially
help to discover as yet unknown processes that are not described in
the pathway databases \cite{Nacu07}.  False negative interactions
form a limitation for the current model because joint responses of
co-regulated genes can be modeled only when they form a connected
subnetwork.

The need for principled methods for analyzing large-scale collections
of gene expression data is increasing with their
availability. Versatile gene expression atlases contain valuable
information about shared and unique mechanisms between disparate
conditions which is not available in smaller and more specific
experiments \cite{Lage08, Scherf2000}. For example, \cite{Lamb06}
demonstrated that large-scale screening of cell lines under diverse
conditions can enhance the finding of therapeutic targets. Our model
is directly applicable in similar exploratory tasks, providing tools
for organism-wide analysis of transcriptional activity in normal human
tissues \cite{Roth06, Su04}, cancer, and other diseases
\cite{Kilpinen08, Lukk10} in a genome- and organism-wide
scale. Similar collections are available for several model
organisms including mouse \cite{Su04}, yeast \cite{Granovskaia10},
and plants \cite{Schmid05}. A key advantage of our approach compared
to methods that perform targeted comparisons between predefined
conditions \cite{Ideker02, Sanguinetti08} is that it allows
systematic organism-wide investigation when the responses and the
associated conditions are unknown.  The motivation is similar to SAMBA
and other biclustering approaches that detect groups of genes that
show coordinated respose in a subset of conditions \cite{Madeira04},
but the network ties the findings more tightly to cell-biological
processes in our model. This can focus the analysis and improve
interpretability. Since the nonparametric mixture model adjusts model
complexity with sample size, our algorithm is potentially applicable
also in smaller and more targeted data sets. For example, it could
potentially advance disease subtype discovery by revealing
differential network activation in subsets of patients.

Many large-scale collections are continuously updated with new
measurements. Our algorithm provides no integration technique for new
experiments yet; on-line extensions that could directly integrate data
from new experiments provide an interesting topic for further study.
Another potential extension would be a fully-Bayesian treatment that
would provide confidence intervals, removing the need to assess
significance of the results in a separate step. While our model
provides a model-based criterion for detecting the responses without
prior knowledge of the activating conditions, the statistical
significance of the findings has to be verified in further
experiments. The majority of the responses in our experiments could be
verified in an independent data set.  Other potential extensions
include adding more structure to address the directionality, relevance
and probabilities of the interactions. Not all cell-biological
processes have clear manifestations at transcriptome level. Hence
information of transcript and interaction types, as in SPIA, could
potentially help to improve the sensitivity of our approach. We could
also seek to loosen the constraints imposed by the prior
network. However, such extensions would come with an increased
computational cost. The simple and efficient implementation is a key
advantage.

NetResponse is closely related to subspace clustering methods such as
agglomerative independent variable component analysis (AIVGA;
\cite{Honkela08}). However, AIVGA and other model-based feature
selection techniques \cite{Law04, Roth04} consider all potential
connections between the features, which leads to a more limited
scalability. Finding a global optimum in our model would require
exhaustive combinatorial search over all potential subnetworks.  Since
the complexity depends on the topology of the network, finding a
general formulation for the model complexity is problematic. The
number of potential solutions grows faster than exponentially with the
number of features (genes) and links between them, making exhaustive
search in genome-scale interaction networks infeasible. Approximative
solutions are needed, and are often sufficient in practice. A
combination of techniques is used to achieve an efficient algorithm
compared to the model complexity. First, we focus the analysis on
those parts of the data that are supported by known interactions. This
increases modeling power and considerably limits the search
space. Second, the agglomerative scheme finds an approximative
solution where at each step the subnetwork pair that leads to the
highest improvement in cost function is merged. This finds a solution
relatively fast compared to the complexity of the task. Note that the
order in which the subnetworks become merged may affect the
solution. Finally, the variational implementation considerably speeds
up mixture modeling \cite{Kurihara07nips}. The running time of our
application was 248 min on a standard desktop computer (Intel 2.83GHz;
Supplementary Fig. 5).

Investigation of a human pathway interaction network revealed
condition-specific regulation in the network, that is, groups of
interacting genes whose joint response differs between physiological
conditions. This highlights the condition-dependent nature of network
activation, and emphasizes an important shortcoming in the current
gene set-based testing methods \cite{Nam08}: simply measuring gene
set 'activation' is often not sufficient; it is also crucial to
characterize {\it how} the expression changes, and in which
conditions. Organism-wide modeling can provide quantitative
information about these connections.

\section*{Conclusions}

We have introduced and validated a general-purpose algorithm for
global identification and characterization of transcriptional
responses in genome-scale interaction networks across diverse
physiological conditions. An organism-wide analysis of a human pathway
interaction network validates the model, and provides a global view on
cell-biological network activation. The results reveal shared and
unique mechanisms between physiological conditions, and potentially
help to formulate novel hypotheses of gene function in previously
unexplored contexts.

\section*{Funding}
This work was supported by the Academy of Finland [207467] and the IST
Programme of the European Community, under the PASCAL 2 Network of
Excellence [ICT-216886]. LL and SK belong to the Finnish CoE on
Adaptive Informatics Research Centre of Academy of Finland, and to the
Helsinki Institute for Information Technology HIIT.

\bibliographystyle{abbrv}

\newpage

\onecolumn

\section*{Supplementary Figures}


\begin{figure*}[h!]
\begin{center}

\begin{tabular}{cc}
{\bf A}&\rotatebox{270}{\resizebox{!}{6.5cm}{\includegraphics{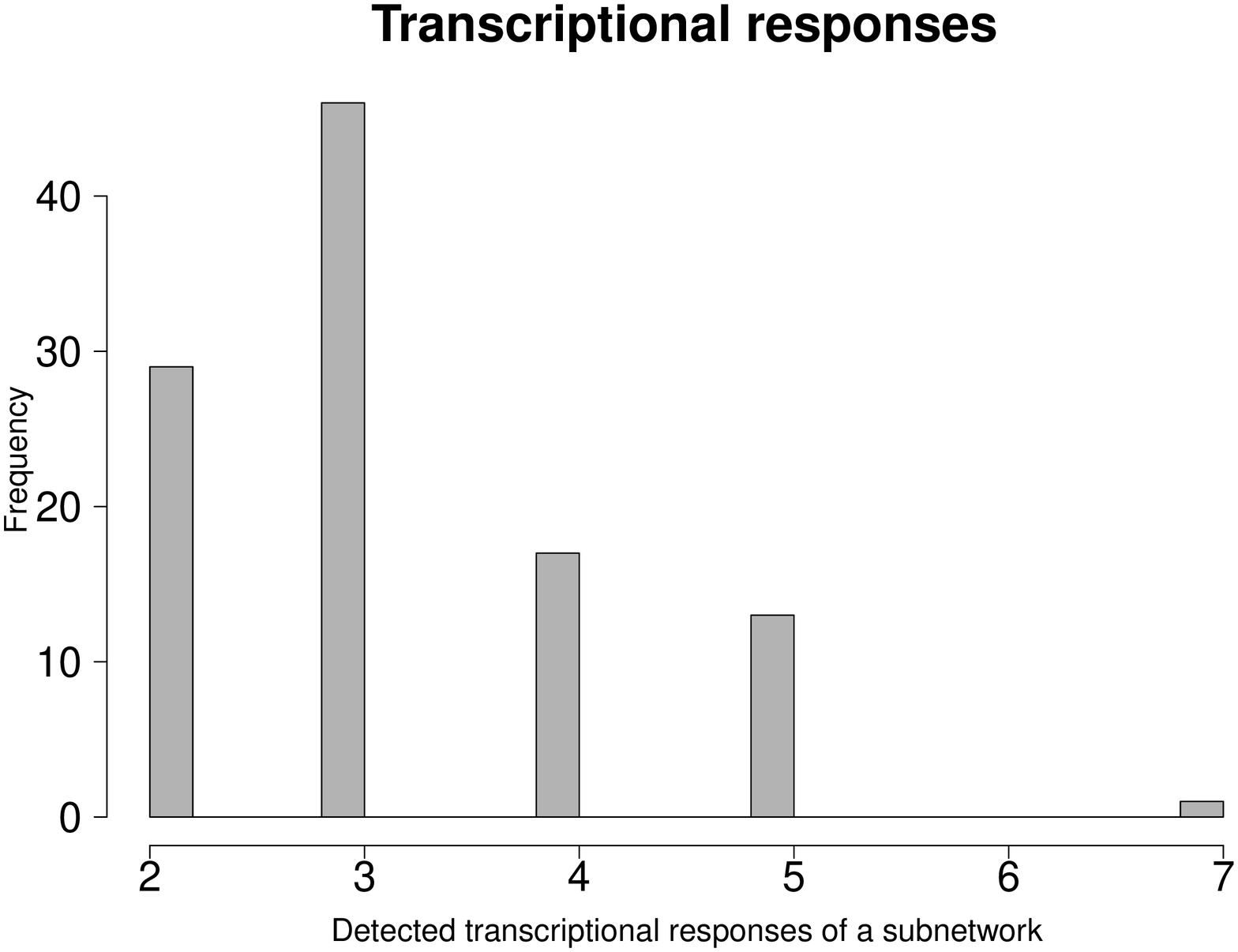}}}\\
{\bf B}&\rotatebox{270}{\resizebox{!}{6.5cm}{\includegraphics{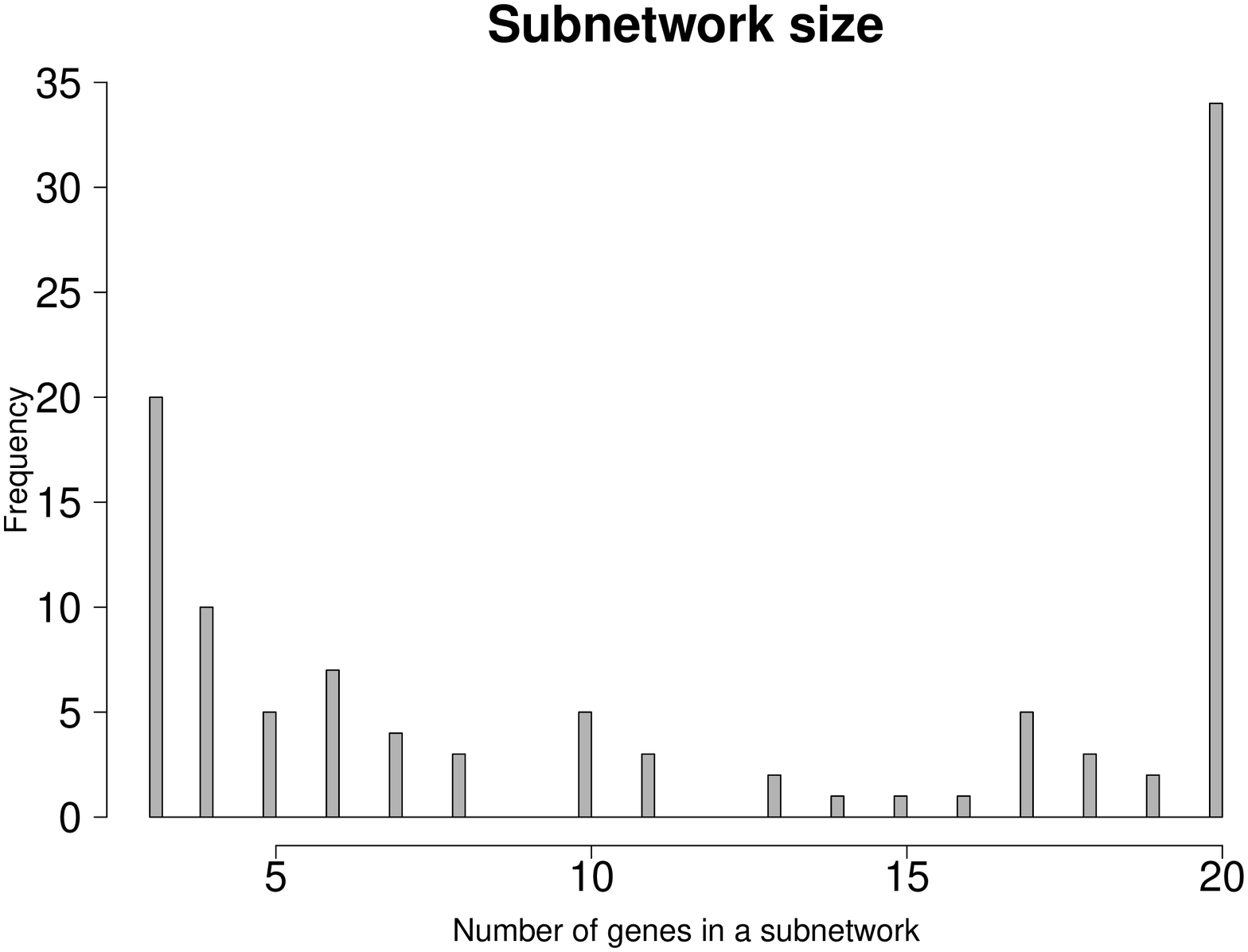}}}\\
{\bf C}&\rotatebox{270}{\resizebox{!}{6.5cm}{\includegraphics{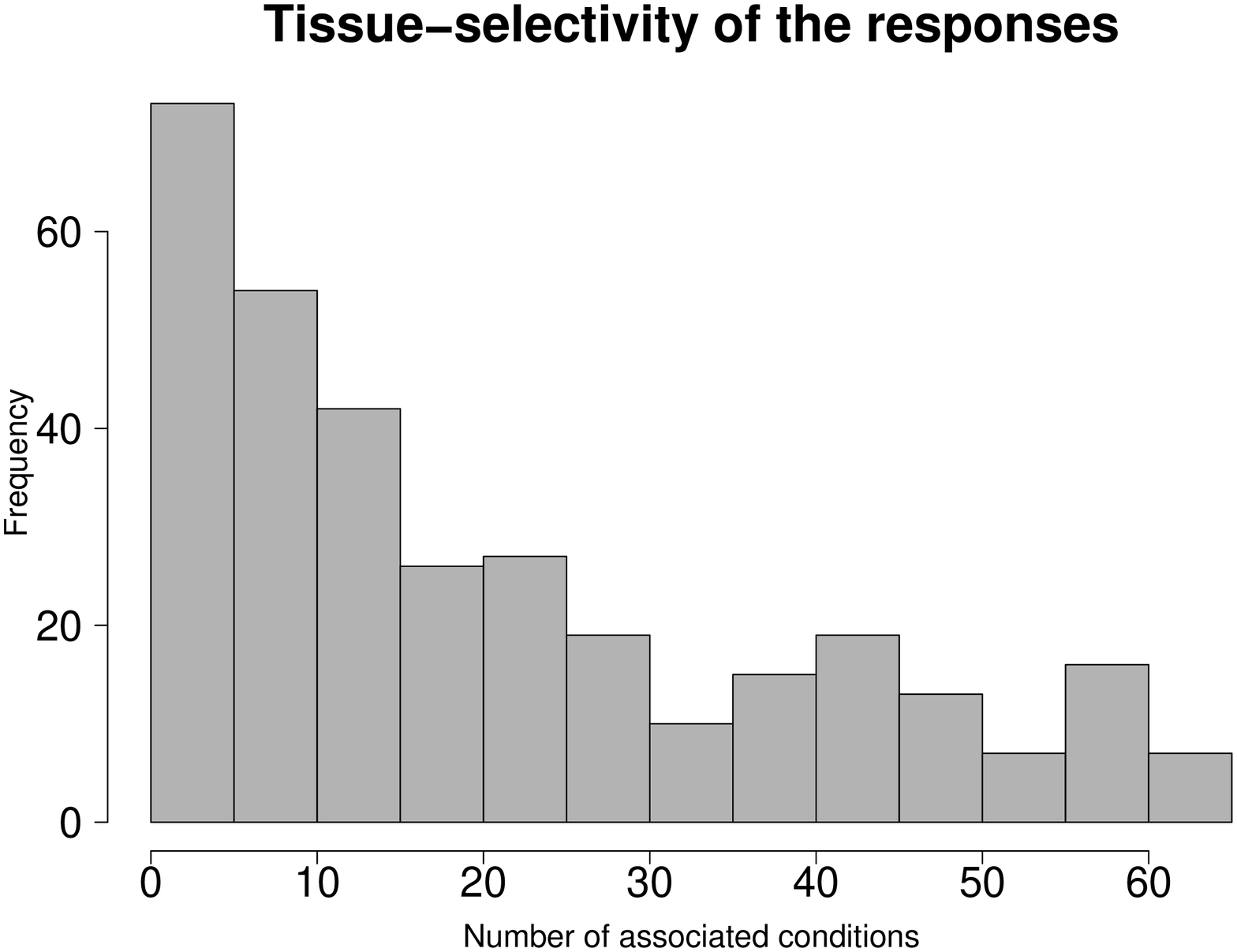}}}
\end{tabular}
\end{center}
\caption{{\bf Supplementary Fig 1}. Histograms of model statistics: {\bf A} Number of
  transcriptional responses in the subnetworks detected by
  NetResponse. {\bf B} Subnetwork size. {\bf C} Number of physiolocial
  conditions associated with each response.}
\end{figure*}



\begin{figure*}[h]
\begin{center}
\rotatebox{270}{\resizebox{!}{14cm}{\includegraphics{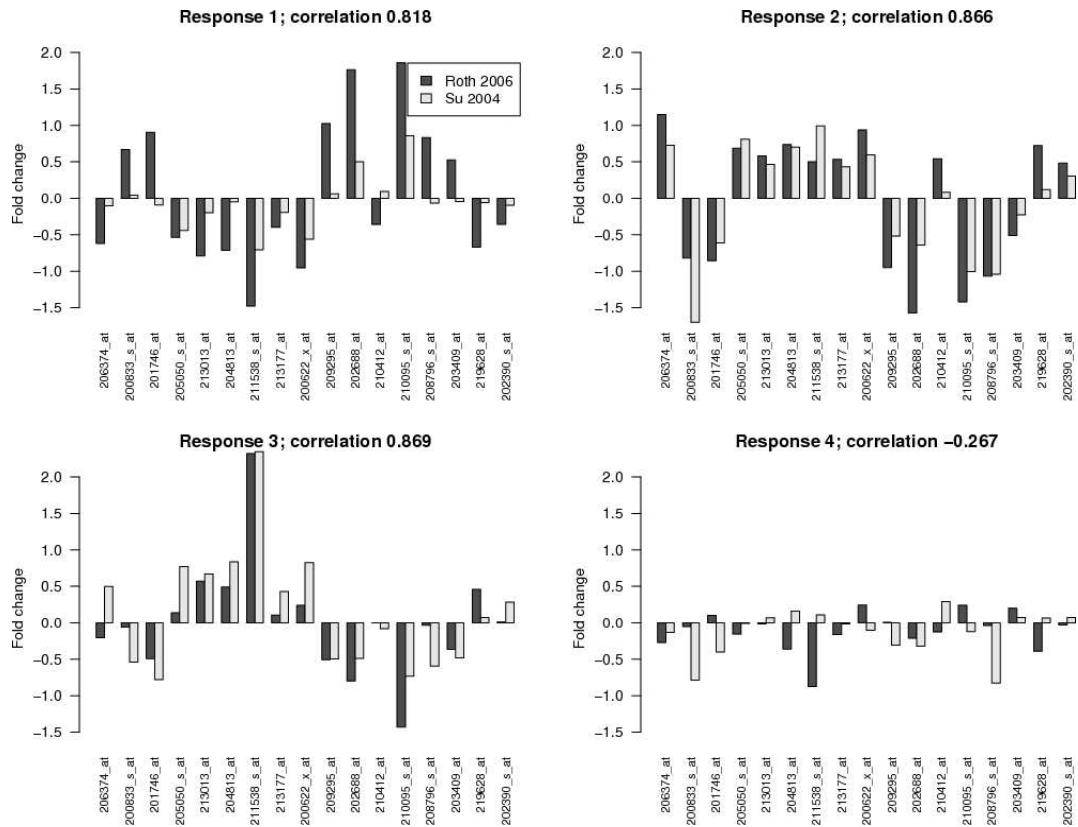}}}
\end{center}
\caption{{\bf Supplementary Fig 2} Reproducibility of transcriptional responses of the
  subnetwork of Figure 1 (main text) in independent validation
  data. {\it Correlation:} Qualitatively similar responses are
  observed in the validation data (Pearson correlation \(> 0.8\)),
  except for the fourth response (correlation -0.27).  Differential
  expression with respect to the mean level of each gene is used in
  the comparisons. This removes overoptimistic bias in the
  correlations caused by the systematic differences in the expression
  levels of the genes.  {\it Significance:} Each response is
  associated with a subset of conditions.  The differences between the
  corresponding conditions are statistically significant (\(p <
  0.01\); GlobalTest) in the validation data for each pairwise
  comparison between the predicted four groups of conditions. The
  investigated gene expression atlas (Roth et al., 2006) and the
  validation data (Su et al., 2004) have been measured on different
  array platforms (HG-U133Plus2 and HG-U133A, respectively). Gene
  expression levels are here shown for the 17 (out of 20) probesets
  that are available on both platforms.}
\end{figure*}



\begin{center}
\begin{figure*}[h!]
\rotatebox{180}{\resizebox{!}{18cm}{\includegraphics{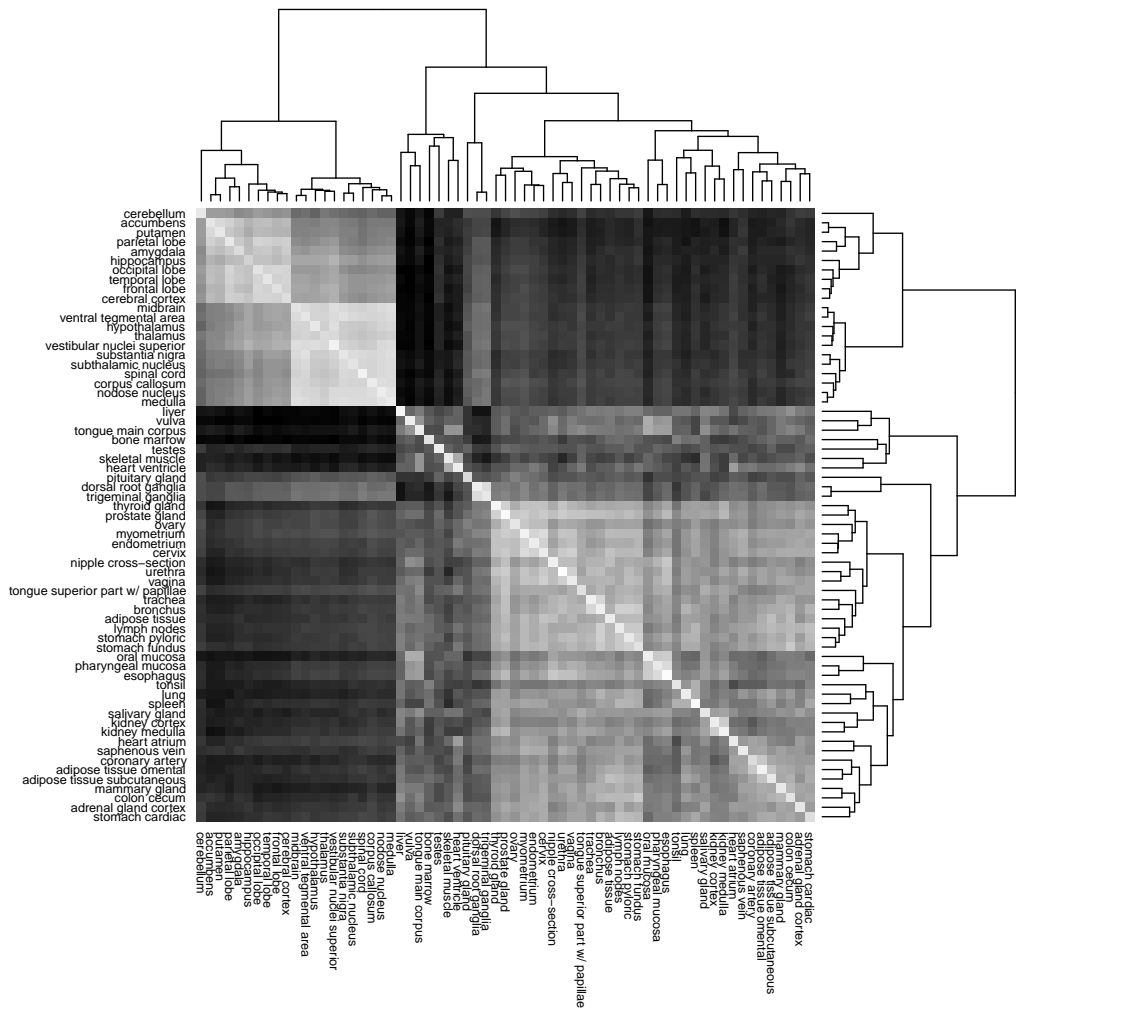}}}
\vspace{-5mm}
\caption{{\bf Supplementary Fig 3} Tissue connectome based on the detected transcriptional
  responses of the human pathway interaction network.  For each pair
  of tissues the overall probability of shared transcriptional
  response across the network is shown (black: \(P = 0\); white: \(P =
  1\); see main text for details). This gives a probabilistic measure
  of tissue similarity based on network activation. The rows and
  columns are ordered with hierarchical clustering to highlight the
  relatedness between physiological conditions.}
\end{figure*}
\end{center}



\begin{figure*}[h]
\begin{center}
\begin{tabular}{cc}
{\bf A}&\rotatebox{270}{\resizebox{!}{10cm}{\includegraphics{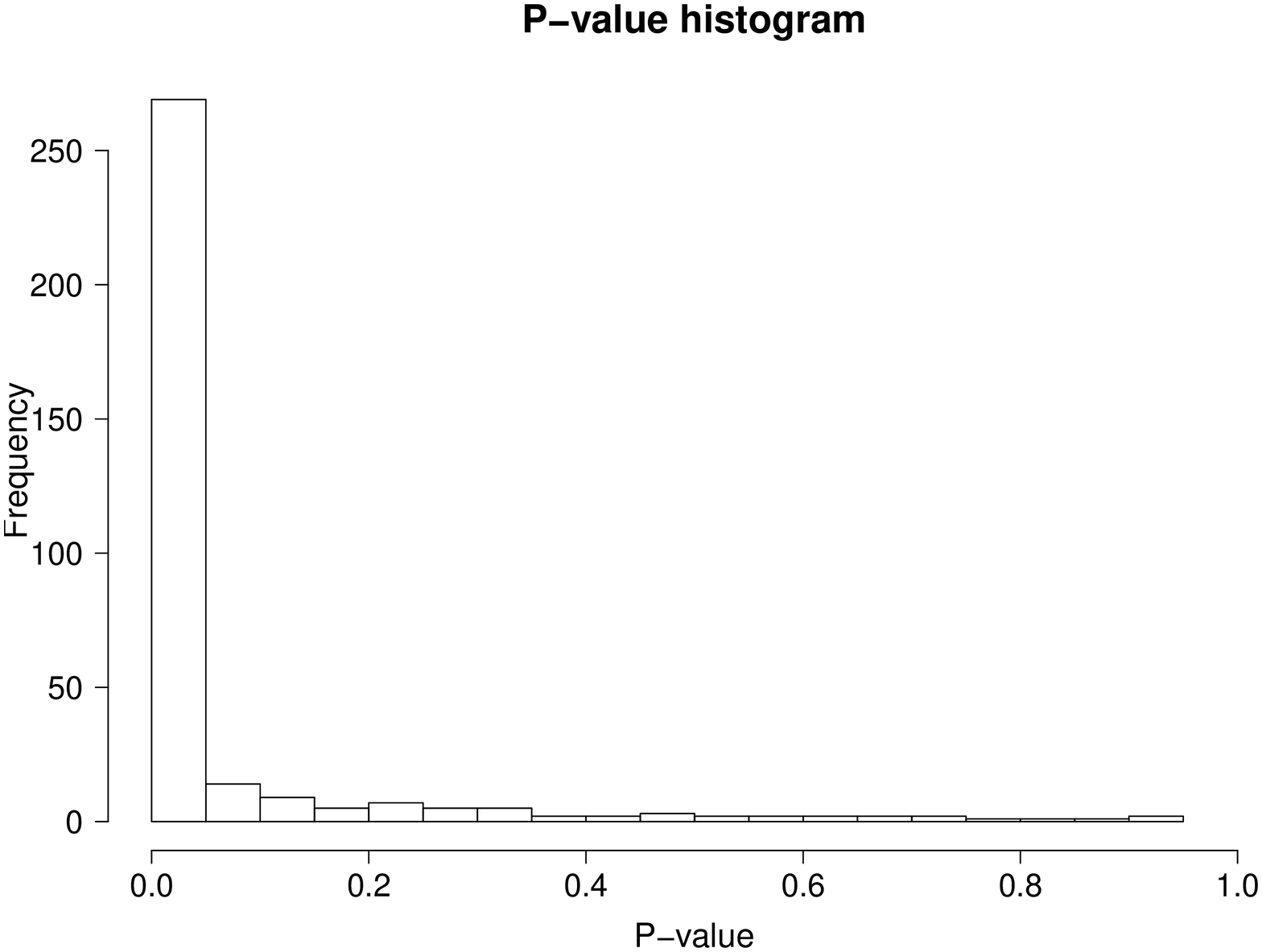}}}\\
{\bf B}&\rotatebox{270}{\resizebox{!}{10cm}{\includegraphics{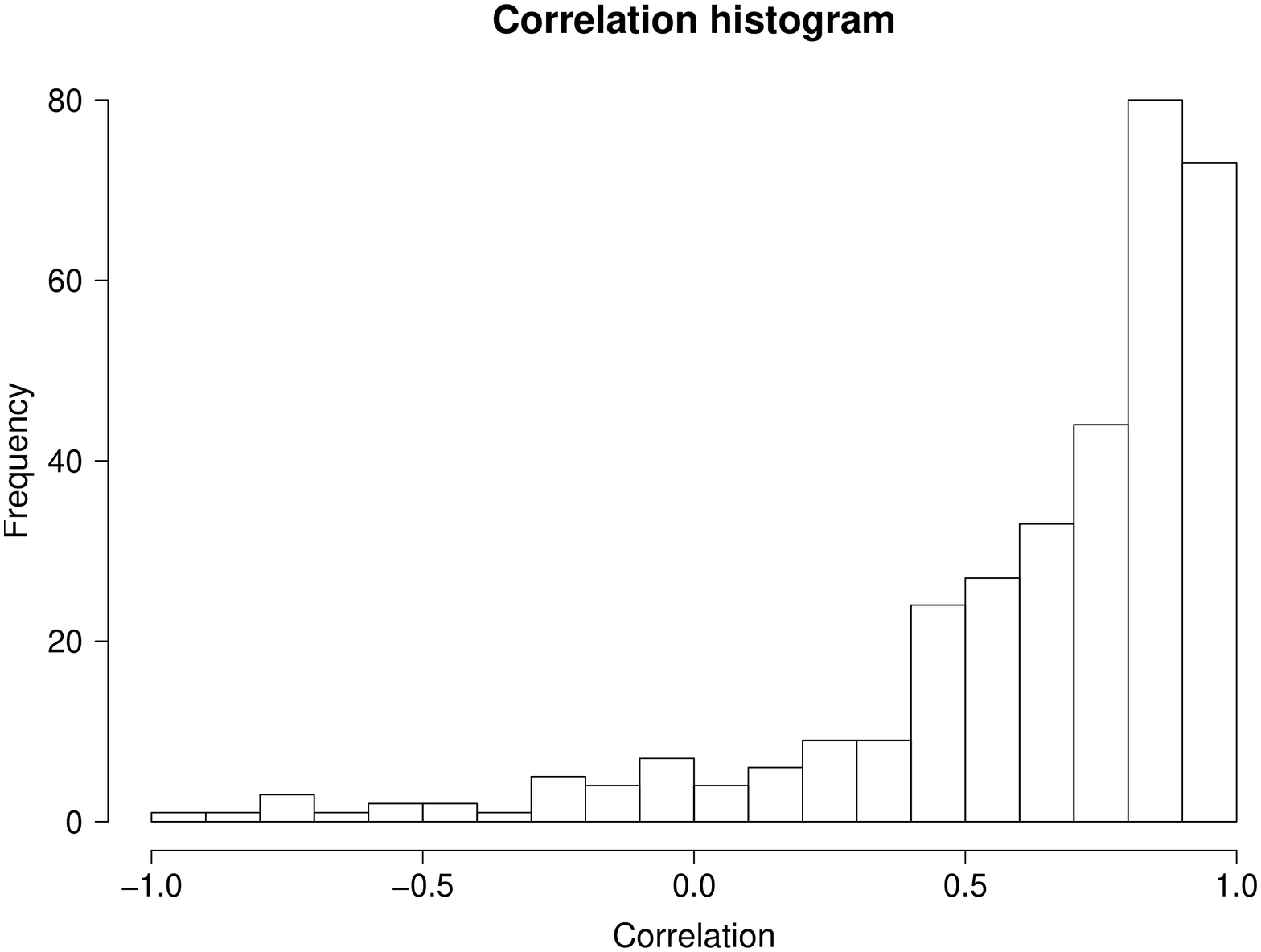}}}
\end{tabular}
\end{center}
\caption{{\bf Supplementary Fig 4} Reproducibility of the detected transcriptional responses in
  the independent Su et al., 2004 validation data in terms of
  significance and correlation. {\bf A} Significance of differential
  expression between each pair of associated conditions for predicted
  responses in the validation data. 80\% of the predicted differences
  between the conditions were verified in the validation data with \(p
  < 0.05\) (GlobalTest). We tested only the responses where
  corresponding conditions were available in the validation data (81\%
  of the responses). {\bf B} Correlation between the detected
  responses in the investigated data set and the corresponding
  conditions in the validation data. Differential expression with
  respect to the mean level of each gene was used in the
  comparisons. This removes the potential bias in the correlations
  caused by the systematic differences in the expression levels of the
  genes.}
\end{figure*}



\begin{figure*}[h]
\begin{center}
\rotatebox{270}{\resizebox{!}{10cm}{\includegraphics{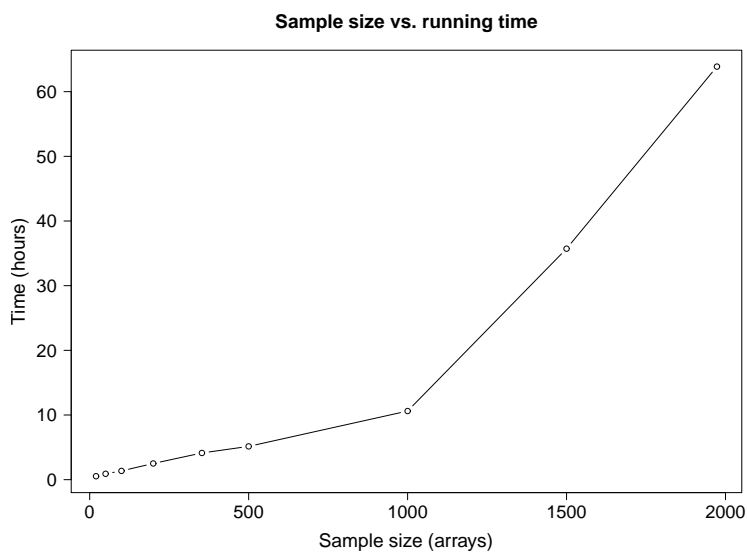}}}
\end{center}
\caption{{\bf Supplementary Fig 5} Running time for data sets of different sizes on the pathway
  network described in the main text. The running time for the GSE3526
  data set investigated in the main text was 248 minutes (i.e. 4.1
  hours). Computation time increases superlinearly with sample size
  from 33 minutes with 20 samples to 64 hours with 1977 samples. Model
  fitting in the algorithm can be parallelized, which will make the
  model scalable to larger data sets in standard multi-core desktop
  computers. The running time depends also on the size and
  connectivity of the network. Our investigated network represents a
  standard pathway network used in current organism-wide studies. The
  network has a median of 5 and a maximum of 105 direct interaction
  partners per gene. This reduces the search space considerably
  compared to models that would consider all potential interactions
  between the 1800 network genes. To investigate time consumption we
  have selected random subsets of various sizes (20, 50, 100, 200, and
  353 samples) from the GSE3526 data, described in the main text and
  having 353 arrays in total. The data sets with 500 and more (1000,
  1500, 1973) samples were obtained by picking random subsets from the
  GSE2109 data set, which has 1973 arrays in total (downloaded
  30.5.2008 from http://www.ncbi.nlm.nih.gov/geo/). Both data sets
  were preprocessed as described in Section 2.3 in the main text.}
\end{figure*}


\begin{table}[h]
\small{
\begin{center}
\begin{tabular}{llllll}
  \hline\\
 & NetRespose & NetResponse& MATISSE+ &MATISSE+  & SAMBA \\ 
 &  & (shuffled) &         & (shuffled) &  \\ 
  \hline\\
Reproducibility (corr.) & 0.76 & 0.14 & 0.64 & 0.60 & 0.68 \\ 
Reproducibility (signif.) & 0.80 & 0.43 & 0.72 & 0.71 & 0.63 \\ 
Fraction of responses & 0.81 & 0.34 & 0.79 & 0.80 & 0.89 \\ 
with validation data \\
Physiol. coh. (NMI) & 0.50 & 0.22 & 0.49 & 0.41 & 0.46 \\ 
Physiol. coh. (signif.) & \(<10^{-4}\) & \(0.65\) & \(<10^{-4}\) & \(<10^{-2}\) & \(<10^{-4}\) \\ 
Fraction of data  & 0.68 & 0.72 & 0.45 & 0.40 & 0.45 \\ 
assigned to subnetworks\\
\hline
\end{tabular}
\end{center}
}
\caption{{\bf Supplementary Table 1} Comparison statistics.
  {\it Reproducibility (correlation):} Median correlation between the
  detected responses and the corresponding conditions in the
  validation data.
  {\it Reproducibility (significance):} Fraction of transcriptional responses that were reproducible in the validation data (GlobalTest \(p<0.05\)). The results are shown for the responses where corresponding conditions in the validation data were available. Significance of differential
  expression was calculated for each pairwise comparison between the associated conditions of the predicted responses in the validation data.
  {\it Transcriptional responses with validation data:} Fraction of transcriptional responses for which corresponding samples were available for testing in the validation data.
  {\it Physiological coherence (NMI):} Normalized mutual information between the detected transcriptional responses and sample labels (physiological conditions). A higher NMI indicates stronger association between the detected responses and physiological conditions. The differences between NetResponse, MATISSE+, and SAMBA are not significant.
  {\it Physiological coherence (significance):} Significance of the physiological coherence (NMI) compared to the expectation based on randomly labeled samples (Wilcoxon test p-value).
  {\it Fraction of data assigned to subnetworks:} Fraction of genes participating in the detected responses.}
\end{table}

\end{document}